\documentclass[doublecol]{epl2}
\usepackage{amssymb,amsmath}

\title{Measurement of the Aharonov-Casher geometric phase with a separated-arm atom interferometer}

\author{J.~Gillot, S.~Lepoutre, A.~Gauguet, J.~Vigu\'e and M.~B\"uchner}
\institute{ Laboratoire Collisions Agr\'egats R\'eactivit\'e-IRSAMC \\
 Universit\'e de Toulouse-UPS and CNRS UMR 5589, Toulouse, France}

\pacs{03.75.Dg}{Atom and neutron interferometry}
\pacs{03.65.Vf}{Phases: geometric; dynamic or topological}
\pacs{39.20.+q}{Atom interferometry techniques}

\date{\today}

\abstract{
In this letter, we report a measurement of the Aharonov-Casher (AC) geometric phase with our lithium atom interferometer. The AC phase appears when a particle carrying a magnetic dipole propagates in a transverse electric field. The first measurement of the AC phase was done with a neutron interferometer in 1989 by A. Cimmino \textit{et al.} (Phys. Rev. Lett. \textbf{63}, 380, 1989)  and all the following experiments were done with Ramsey or Ramsey-Bord\'e interferometers with molecules or atoms. In our experiment, we use lithium atoms pumped in a single hyperfine-Zeeman sublevel and we measure the AC-phase by applying opposite electric fields on the two interferometer arms. Our measurements are in good agreement with the expected theoretical values and they prove that this phase is independent of the atom velocity.}

\begin{document}

\maketitle

In 1984, Y.~Aharonov, and A.~Casher \cite{AharonovPRL84} discovered the geometric phase which is called by their name. This phase appears when a particle with a  magnetic dipole interacts with an electric field perpendicular to both the particle velocity and to the magnetic dipole. This phase had also been discussed by J.~Anandan \cite{AnandanPRL82}, who did not remark its unusual properties.  The Aharonov-Casher (AC) phase is the second example of a geometric phase, after the Aharonov-Bohm phase \cite{AharonovPR59}. A third geometric phase predicted in 1993/1994 by X.-G.~He, B.H.J.~McKellar \cite{HeMcKellar93} and by M.~Wilkens \cite{Wilkens94} is now named the He-McKellar-Wilkens (HMW) phase, and we have recently measured this phase \cite{Lepoutre12,GillotPRL13}. This phase appears, if an electric dipole travels in a magnetic field $\mathbf{B}$ and if the scalar triple product of the dipole, $\mathbf{B}$ and the velocity vectors is not zero. All these phases  belong to the general class of geometric phases discussed by M.V.~Berry in 1984 \cite{BerryPRS84,Shapere89} and they are very interesting because they strongly differ from dynamical phases: geometric phases modify the wave propagation in the absence of any force on the particle; they are independent of the modulus of the particle velocity but they change sign if the velocity is reversed.

In the present letter, we describe measurements of the AC phase shift using a separated-arm $^7$Li atom interferometer using Bragg diffraction on laser standing waves.

The internal quantum state of the atom is the same in the two interferometer arms and we apply opposite electric fields and a common magnetic field on the two interferometer arms. The atom fringes are phase shifted by both the AC and the HMW phases. The AC phase is proportional to the atom magnetic dipole moment, thus depends on the magnetic sublevel, while the HMW phase is independent.
By combining measurements made with the $^7$Li atoms pumped in $F=2$ either in $m_F=+2$ or in $m_F=-2$,  we can extract both phases. In the following, we focus on the AC phase measurements. As explained below, this is the first AC phase measurement of this type and  the sensitivity of our atom interferometer has enabled us to test the velocity dependence of this phase.

\begin{largetable}
\caption{Previous measurements of the AC phase. $\mu$ is the magnetic dipole moment and $\mu_B$ the Bohr magneton. $E_{max}$ is the maximum electric field strength, $\varphi_{AC,max}$ is the maximum AC phase and $v$ is the particle velocity. Ref. \cite{ZeiskeAPB95} and ref. \cite{Yanagimachi02} did not probe the same quantum superposition configuration, which explains the larger ratio $\varphi_{AC,max}$/$E_{max}$ of \cite{Yanagimachi02}. \label{table1}}

\begin{center}
 \begin{tabular}{|lccccc|}
    \hline
    Species                    &  $\mu/\mu_B$         &  $E_{max}$  &$\varphi_{AC,max}$& error  & $v$   \\
                               &                      &     (MV/m)    &       (mrad)       &    (\%)     &  (m/s)   \\ \hline
    n \cite{CimminoPRL89}& $1.0\times 10^{-3}$ &      $30$   &  $ 2.19 $        &    $24$   & $2680$          \\
    TlF \cite{SangsterPRL93}   & $1.4\times 10^{-3}$ &      $3$    &  $ 2.22 $        &    $4$    & $220-340$ \\
    TlF \cite{SangsterPRA95}   & $1.4\times 10^{-3}$ &      $2$    &  $ 2.42 $        &    $2$    & $188-366$ \\
    $^{85}$Rb \cite{GorlitzPRA95}     & $1/3$              &      $0.9$  & $  150 $        &    $1.4$  & $300-650$ \\
    $^{40}$Ca \cite{ZeiskeAPB95}      & $3/2$               &      $1$    & $  35  $        &    $2.2$  & $643-698$ \\
    $^{40}$Ca \cite{Yanagimachi02}    & $3/2$               & 4.4         &        $ 314 $      & $2.9$  & $650, 810$ \\ \hline
\end{tabular}
\end{center}
\end{largetable}

Since its theoretical discovery, the AC phase has been tested in five different experiments \cite{CimminoPRL89,SangsterPRL93,SangsterPRA95,GorlitzPRA95,ZeiskeAPB95,Yanagimachi02}. We recall in table \ref{table1} the main parameters of these experiments. In the first experiment done in 1989 by A.~Cimmino \textit{et al.} \cite{CimminoPRL89} with a neutron interferometer, supplementary phase shifts of magnetic and gravitational origins were needed in order to measure a spin-dependent phase with unpolarized neutrons. The four following experiments were based on Ramsey or Ramsey-Bord\'e interferometry with atoms or molecules. In these experiments, the atom or molecule propagates in a quantum superposition of internal states with different magnetic moments and the AC phase shift appears directly as a shift of the fringe signal.
We do not use such quantum superpositions in our experiment: the atom is in the same internal quantum state in the two interferometer arms and we deduce the AC phase by comparing the fringe phase shift measured during experiments made with different atom quantum states.

The AC phase is given by:

\begin{equation}
\varphi_{AC} = - \frac{1}{\hbar c^2} \oint \left[\mathbf{E} \left( \mathbf{r} \right) \times \boldsymbol{\mu} \right] \cdot d \mathbf{r}
\label{PhiAC}
\end{equation}

\noindent where $\boldsymbol{\mu}$ is the particle magnetic dipole and $\mathbf{E}$ the electric field. A.G.~Klein
\cite{KleinPhysica86} remarked that, at first order in $v/c$,  the AC phase can be interpreted as the interaction
of the magnetic moment $\boldsymbol{\mu}$ with the motional magnetic field $\mathbf{B}_{mot}\approx - \left(\mathbf{v} \times
\mathbf{E}\right) / c^2$ seen by the particle in its rest frame moving with the velocity $\mathbf{v}$.  In the presence of
 a magnetic field $\mathbf{B}$, the  particle interacts with the total field $\mathbf{B}+ \mathbf{B}_{mot}$ and this interaction induces a  phase shift $\varphi_{Z+AC}$ due to the sum of the Zeeman and AC phase effects:

\begin{equation}
\varphi_{Z+AC}(F,m_F) = - \frac{1}{\hbar} \oint E_{F,m_F}(|\mathbf{B}+\mathbf{B}_{mot}|)dt
\label{Emag}
\end{equation}

\noindent where $E_{F,m_F}(|\mathbf{B}+\mathbf{B}_{mot}|)$  is the energy of the $(F,m_F)$ sublevel in the presence of the field. We measure the AC phase by measuring the variation of $\varphi_{Z+AC}$ due to the presence of $\mathbf{B}_{mot}$. In our experiment, $B_{mot} \leq 10^{-8}$ T is always considerably smaller than $ B \geq 10^{-5}$ T, and the AC phase can be written as:

\begin{eqnarray}
\varphi_{AC}(F,m_F)= - \frac{1}{\hbar} \oint \frac{\partial E_{F,m_F}}{\partial B}   \left( \mathbf{B}_{mot} \cdot \hat{\mathbf{e}}_{B} \right) dt
\label{Emagappro}
\end{eqnarray}

\noindent where $ \hat{\mathbf{e}}_{B}$ is a unit vector parallel to $\mathbf{B}$, $ \hat{\mathbf{e}}_{B}=\mathbf{B}/B $: only the component of $\mathbf{B}_{mot}$ parallel to $\mathbf{B}$ contributes to the AC phase.

\begin{figure}[ht]\begin{center}
\includegraphics[width= 8 cm]{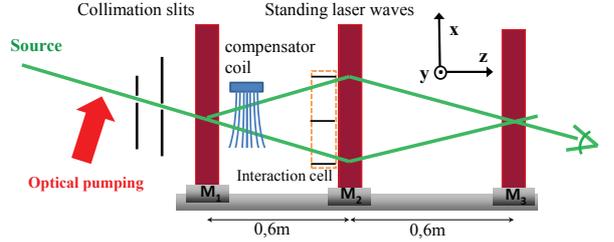}

\caption{  (Color online) Schematic top-view of our atom interferometer (not to scale): a supersonic lithium is optically pumped, collimated  and crosses three laser standing waves, which diffract the atom waves in the Bragg regime. A coil, mounted at mid-distance between the first standing waves produces a compensating magnetic field gradient. The interaction cell is mounted just before the second laser standing wave (see Figure \ref{fig2} for details.)  \label{fig1}}
\end{center}
\end{figure}

Our Mach-Zehnder atom interferometer has been described in detail \cite{MiffreEPJD05} and it is schematically represented in figure \ref{fig1}. The atomic beam is produced by a supersonic expansion of natural lithium seeded in a large excess of a noble gas which fixes the mean beam velocity $v_m$ of the lithium atoms: $v_m$  scales like $1/\sqrt{M}$, where $M$ is the noble gas atomic mass (see table \ref{VI0}). The beam velocity distribution is well described by a Gaussian \cite{ToenniesJCP77,Miller88} with a $1/e$ half-width equal to $v_m/S_{\|}$ where $S_{\|}$ is the parallel speed ratio. $S_{\|}$ depends on the source parameters (nozzle diameter, pressure, temperature, carrier gas), {with typical values $S_{\|}= 6-8$ in our experiment}.

The lithium beam is optically pumped and collimated. Then, it crosses three laser standing waves which diffract the atom wave in the Bragg regime: first-order diffraction is used to split, reflect and recombine the atomic waves. The laser used to produce the standing waves is a single frequency dye laser. Its wavelength $\lambda_L$ is chosen on the blue side of the $^2$S$_{1/2}\rightarrow ^2$P$_{3/2}$ transition of $^7$Li at $671$ nm: this choice and the natural abundance of $^7$Li ($92.5$\%) explain the fact, that only $^7$Li contributes to the interferometer signal \cite{MiffreEPJD05,JacqueyEPL07}. This signal can be written as:

\begin{equation} \label{I0V}
 I=I_0 \left[1+ {\mathcal V} \cos{(\varphi_d + \varphi_p)} \right]
\end{equation}

\noindent where $I_0$ is the mean signal intensity and ${\mathcal V}$ the fringe visibility. $\varphi_d$ is the diffraction phase, $\varphi_d= 4 \pi (x_1-2 x_2+x_3)/\lambda_L$ with $x_i$ the position of the mirror $M_i$ (i=1,2,3), and $\varphi_p$ the phase {\t due to various} perturbations. The  phase sensitivity, i.e. the minimum detectable phase for an one-second data recording is proportional to $\varphi_{min}=1/({\cal V}\sqrt{I_0})$. Table \ref{VI0} summarizes the typical values of $I_0$, ${\mathcal V}$ and $\varphi_{min}$ as a function of the carrier gas used in our experiment.

\begin{table}[ht!]
\caption{The mean lithium beam velocity $v_m$, the mean signal intensity $I_0$, the fringe visibility ${\mathcal V}$ and the phase sensitivity $\varphi_{min}$ measured with the three carrier gases. $I_0$ is large with  neon and argon but considerably smaller with krypton, while ${\mathcal V}$ has opposite variations. $\varphi_{min}$  has comparable values with neon and argon and is about half as much with krypton. \label{VI0}}
\begin{center}
\begin{tabular}{c|c|c|c|c}
    carrier gas            & $v_m$             & $I_0$       & ${\cal V}$ &  $\varphi_{min}$        \\
   $M$ (a.m.u.) & (m/s)               &($10^3$ c/s) &  \%        &  mrad/$\sqrt{{\mbox Hz}}$ \\ \hline
    Ne:    20.2     &  $1520 \pm 38$    &  56         &  60        & 7.0                     \\
    Ar:    39.9     &  $1062 \pm 20$    &  33         &  75        & 7.3                     \\
    Kr:    83.8     &  $ 744 \pm 18$    &  7          &  80        & 14.9                    \\
\end{tabular}
\end{center}
\end{table}

The lithium de Broglie wavelength is $\lambda_{dB}=5.7 \times 10^{-8}/v_m $, with $\lambda_{dB}$ in m and $v_m$ in m/s, and the first order diffraction angle is $\theta = 2\lambda_{dB}/\lambda_L \approx 0.17/v_m$ rad,  for example, with argon as a carrier gas, $\lambda_{dB} \approx 54$ pm and  $\theta \approx 160$ $\mu$rad. The distance between the interferometer arms,  which is maximum at the second laser standing wave, is approximately equal to $143$ (krypton), $100$ (argon) and $70$ $\mu$m (neon), which is sufficient to introduce a septum \cite{EkstromPRA95} between the two interferometer arms.

\begin{figure}
\begin{center}
\includegraphics[width= 8 cm]{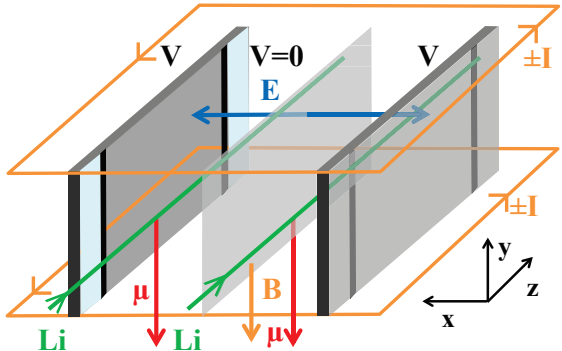}
\caption{  (Color online) Schematic drawing of our interaction cell (not to scale). A double capacitor produces opposite electric fields $\mathbf{E}$ (blue) in the horizontal plane. The high voltage electrodes are surrounded  by grounded guard electrodes (light blue). Two coils (yellow) produce a vertical magnetic field $\mathbf{B}$. The magnetic dipole (red) \boldmath{$\mu$} is parallel or antiparallel to the magnetic field, depending on the pumped level. Each interferometers arm (green) passes through a different capacity.  \label{fig2}}
\end{center}
\end{figure}

We have  built the present experiment to measure the HMW phase shift, using an arrangement inspired by the ideas of H.~Wei \textit{et al.} \cite{WeiPRL95}, with opposite electric fields on the  two interferometer arms and a common homogeneous magnetic field \cite{Lepoutre12,LepoutreHMWII13}. Figure \ref{fig2} shows the interaction cell. It consists of a double plane capacitor used to produce the needed electric fields and the capacitor assembly is inserted in a support which holds two coils producing the magnetic field.

The two capacitors share the septum as a common, grounded electrode. The electric field vector lies on the horizontal plane, with opposite values on the two interferometer arms and a magnitude of $E_{max} \approx 0.7 $ MV/m for the maximum applied voltage $V=800$ V and it extends over a $48$ mm length. Each high voltage electrode is separated by  $1$ mm-wide gaps from two $5$ mm-long grounded guard electrodes, in order to have well defined fringing fields.

The two coils produce a magnetic field \textbf{B} along the vertical axis and with a current of $I$=40 A, we obtain a field strength of $B_{coil}$=22.4 mT. This magnetic field is needed for the observation of the HMW effect but its presence is also important for the observation of the AC phase,  as discussed in equation (\ref{Emagappro}) and also because the magnetic moment of a hyperfine-Zeeman sublevel is parallel or antiparallel to the local magnetic field. The magnetic field produced in the interaction region has a small gradient along the $x$-direction so that it has slightly different values on the two interferometer arms, which results in a Zeeman phase shift. We have introduced a compensator coil to produce an opposite gradient, thus canceling the Zeeman phase shift. This compensator coil is located at mid-distance between the first and second laser standing waves (see figure \ref{fig1}).

As already mentioned, we optically pump the lithium atoms in one Zeeman-hyperfine ground state sublevel $F=2, m_F=+2$  (or $m_F=-2$) \cite{GillotPO13}. The optical pumping is performed before beam collimation in order to avoid heating the transverse motion by  exchange of photon momenta (radiation pressure). We control the magnetic field in the pumping region by three pairs of square Helmholtz coils. The D1 line of lithium is used because of the larger hyperfine splitting of the $^2P_{1/2}$ state. Two circularly polarized laser beams are tuned to the  $^2S_{1/2}, F=1 \rightarrow$ $^2P_{1/2}, F=2$ and the $^2S_{1/2}, F=2 \rightarrow$ $^2P_{1/2}, F=2$ transitions. The first laser beam empties the $F=1$ level, while the second one pumps the atoms into the $F=2,m_F= \pm 2$ level, depending on the chosen circular polarization and on the magnetic field direction.

As discussed below, we separate the HMW and AC phases by reversing the $m_F$ value which is obtained by reversing the magnetic field in the pumping region. The $m_F$ value measured on an axis parallel to the local magnetic field is conserved along the atom propagation, if the condition for an adiabatic transport is fulfilled and our measurements of the pumping efficiency rule out any Majorana spin-flip of the optically pumped atoms along their propagation. We have characterized the pumping efficiency by an atom interferometric method and we have found that the pumped sublevel has a fraction of the total population near $95\pm 5$\% \cite{GillotPO13}. Our results are recalled in table \ref{POresults}.

\begin{table}[ht]
\caption{The measured population $P(F=2, m_F)$ after optical pumping for different atom velocities $v_m$ \cite{GillotPO13}.   \label{POresults}}
\begin{center}
\begin{tabular}{l|c|r}
    $v_m$ (m/s)             &     $m_F$    &      $P(F=2, m_F)$     \\ \hline
    $744$             & $+2$ & ($96$   $\pm  6$) \% \\
                            & $-2$ & ($93$   $\pm  7$) \% \\
    $1062$            & $+2$ & ($100$  $\pm  13$)\% \\
                            & $-2$ & ($95$   $\pm  11$)\% \\
    $1520$            & $+2$ & ($90$   $\pm  1$) \% \\
                            & $-2$ & ($94$   $\pm  2$) \% \\
\end{tabular}
\end{center}
\end{table}

For the $^2S_{1/2}, F=2, m_F=\pm 2$ sublevels of $^7$Li, $|\mu/\mu_B| \approx 1$. Using eq. (\ref{Emagappro}) and assuming $\mathbf{B}_{mot}$ parallel to ${\mathbf{e}}_{B}$, we predict an AC phase slope $\left|\partial \varphi_{AC}/ \partial V\right|= (8.57  \pm 0.05) \times 10^{-5}  $ rad/V with an uncertainty due to the capacitor geometry. We thus predict $\left|\varphi_{AC}\right|\approx 69$ mrad for the largest voltage $V= 800$ V.

As explained in our papers \cite{Lepoutre12,GillotPRL13,LepoutreHMWII13},  we eliminate interferometer phase drifts by alternating 6 field configurations, $(V,I)$, $(V,0)$, $(0,I)$, $(-V,0)$, $(-V,I)$ and $(0,0)$  during each $20$ second-long fringe scan. Least-square fits are used  to extract the phase $\varphi(V,I,m_F)$ corresponding to each field configuration and to an optical pumping in the $F=2, m_F=+2$ sublevel. We reduce the statistical uncertainty by averaging the results of about $80$ similar fringe scans. Successive experiments are made with opposite values of the current $I$ and of the $m_F$ value. Five effects contribute to the measured phase shifts. There are two related to the residual polarizability and Zeeman phases due to the differences of electric fields and magnetic fields on the two interferometer arms. The third concerns the HMW phase $\varphi_{HMW}(V,I)$ and two others the AC phases $\varphi_{AC}(V,I)$ and $\varphi_{AC}(V,I=0)$. The following combination cancels the residual Stark and Zeeman phases:

\begin{equation} \label{varphiEB}
 \varphi_{EB}(V,I,m_F)= \varphi(V,I)- \varphi(V,0)- \varphi(0,I) + \varphi(0,0)
\end{equation}

\noindent and  $\varphi_{EB}(V,I,m_F)$ is equal to:

\begin{eqnarray} \label{varphiEBbis}
 \varphi_{EB}(V,I,m_F)&=& \varphi_{HMW}(V,I) + \varphi_{AC}(V,I,m_F) \nonumber \\
 &&- \varphi_{AC}(V,I=0,m_F)
\end{eqnarray}

\noindent Following equation (\ref{Emagappro}), $\varphi_{AC}(V,I=0,m_F)$ is sensitive to the laboratory residual magnetic field  $\textbf{B}_{lab}$ when $I=0$. $B_{lab}$ is about 35 $\mu$T and the field is mainly oriented downwards along the $y$-axis.
If we apply a coil current $I \geq 5$A, the coil magnetic field $B_{coil}$ is always larger than $2.8$ mT and $\mathbf{B}=\mathbf{B}_{coil}+\mathbf{B}_{lab}$ is nearly perfectly vertical. As the HMW phase does not depend on $m_F$ while the AC phase changes sign with $m_F$, we extract the contributions due to the AC phase thanks to the linear combination :

\begin{eqnarray} \label{PhiACEXP}
 \varphi_{AC}^{exp}(V,I)&=&  \left[ \varphi_{EB}(V,I,2) -\varphi_{EB}(V,I,-2) \right]/2
\end{eqnarray}

\noindent Following equation (\ref{varphiEBbis}), $ \varphi_{AC}^{exp}(V,I)$ should be equal to the theoretical value:

\begin{eqnarray} \label{PhiACEXP2}
 \varphi_{AC}^{exp}(V,I)=  \varphi_{AC}(V,I,2)- \varphi_{AC}(V,I=0,2)
\end{eqnarray}

\begin{figure}
\begin{center}
\includegraphics[width= 8 cm]{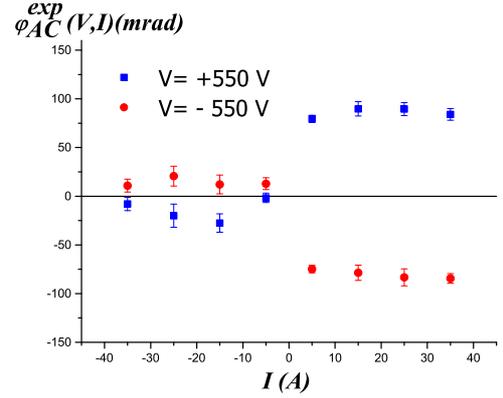}
\caption{  (Color online) The AC phase $\varphi_{AC}^{exp}(V,I)$ defined by equation (\ref{PhiACEXP}) is plotted as a function of the applied current for two voltages $V= \pm 550$V.  The sign $\varphi_{AC}(V,I,m_F)$ is related to the sign of $I$ and its absolute value is larger than $\varphi_{AC}(V,I=0,m_F)$, due to the better verticality of  $\mathbf{B}$, as shown by equation (\ref{Emagappro}).   \label{fig3}}
\end{center}
\end{figure}

Figure \ref{fig3} shows  typical measurements of the AC phase $\varphi^{exp}_{AC}(V,I)$ for two voltages $V=\pm 550$ V as a function of the current $I$. For a given  sense of $I$, $\varphi_{AC}^{exp}(V,I)$ does not vary significantly with the current as expected but $\varphi_{AC}^{exp}(V,I)$ changes sign when we reverse the current $I$ and the magnetic field $\mathbf{B}$. For $I \leq -5$ A, $\mathbf{B}$ is obviously pointing in a direction opposite to the one of $\mathbf{B}_{lab}$ and $\varphi_{AC}(V,I,m_F)$ has a sign opposite to the one of $\varphi_{AC}(V,I=0,m_F)$. As the modulus of $\varphi_{AC}(V,I,m_F)$ is larger than the one of $\varphi_{AC}(V,I=0,m_F)$,  the modulus of $\varphi_{AC}^{exp}(V,I)$ is small but non vanishing when $I \leq -5$ A. For $I \geq 5$ A, the situation is reversed, with $\varphi_{AC}(V,I,m_F)$ and $\varphi_{AC}(V,I=0,m_F)$ having the same sign and adding their contributions to $\varphi_{AC}^{exp}(V,I)$.

\begin{figure}
\begin{center}
\includegraphics[width= 8 cm]{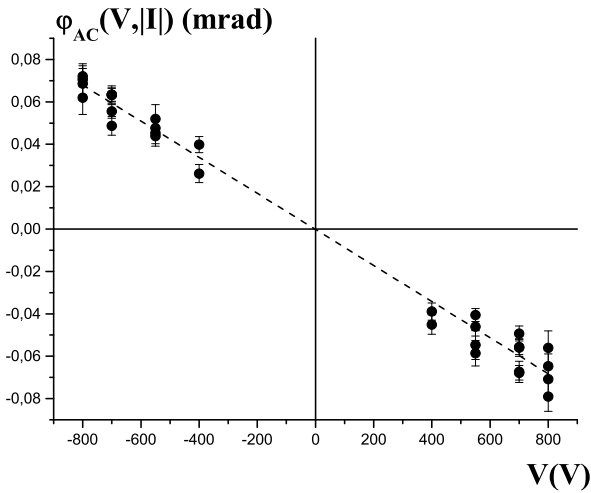}
\caption{  The AC phase $ \varphi_{AC}(V,|I|)$, defined by equation (\ref{ACphaseLC}), is plotted as a function of the applied voltage $V$ for an experiment with a mean lithium velocity $v_m$=1062 m/s.  The data points (black circles) exhibit a linear behaviour and the dashed line represents the best least-square fit. \label{fig4}}
\end{center}
\end{figure}

Using equation (\ref{PhiACEXP2}),  the AC phase in the presence of the applied magnetic field $\mathbf{B}$, $\varphi_{AC}(V,I,2)$, can be extracted by the linear combination:

\begin{equation} \label{ACphaseLC}
 \varphi_{AC}(V,|I|)= \frac{1}{2} \left[\varphi^{exp}_{AC}(V,I)-\varphi^{exp}_{AC}(V,-I) \right]
\end{equation}

\noindent  and $\varphi_{AC}(V,|I|)$ must be equal to $\varphi_{AC}(V,I,2)$.
 Figure \ref{fig4} shows $\varphi_{AC}(V,|I|)$ as a function of $V$  for different values for $|I|$.

\begin{figure}[ht]
\begin{center}
\includegraphics[width= 8 cm]{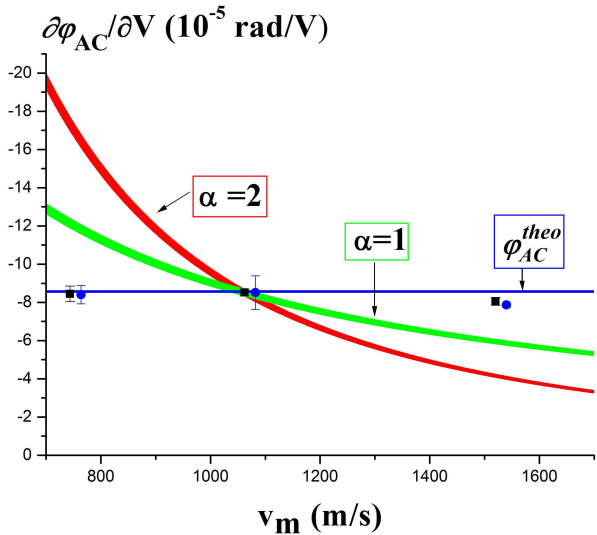}
\caption{  (Color online) Plot of the slope of the AC phase $\partial \varphi_{AC}(V)/\partial V$ as a function of the mean atom velocity $v_m$. The experimental results (black squares) are compared to the theoretical value (blue horizontal band) which assumes a perfect optical pumping. The corrected theoretical values are represented by blue circles slightly displaced at the right of the experimental values for better visibility). The shaded areas represent what would be the phase if, starting from its value for $v_m= 1062$ m/s, the AC phase was varying like $1/v_m^{\alpha}$  with $\alpha=1$ (green) or $\alpha=2$ (red). \label{fig5}}
\end{center}
\end{figure}

The data points plotted with their $1\sigma$ error bars represent a quasi linear behaviour and a least-square fit provides a slope equal to ($-8.51 \pm 0.18$) $\times$ $10^{-5}$ rad/V and a zero-compatible offset ($0.2\pm 1.2$) mrad. We performed similar experiments and data analysis with the three different mean velocities of the lithium beam. The results are collected in Table \ref{table3} and plotted in Figure \ref{fig5}.

\begin{table}
\caption{Our measurements of the AC phase slope $\partial \varphi_{AC}/\partial V$ in $10^{-5}$ rad/V are compared to the theoretical values corrected for imperfect optical pumping (see eq. (\ref{IallMF})).  \label{table3}}
\begin{center}
\begin{tabular}{c|c|c}
                  &  $\partial \varphi_{AC}/\partial V$ & $\partial \varphi^c_{AC}/\partial V$ \\
   $v_m$ (m/s)      &  experiment                         & theory                             \\ \hline
 $1520 \pm 18$    & $-8.05 \pm 0.20 $                   & $-7.86 \pm 0.12 $                  \\
 $1062 \pm 20$    & $-8.51 \pm 0.18 $                   & $-8.51 \pm 0.88 $                  \\
 $ 744 \pm 18$    & $-8.44 \pm 0.41 $                   & $-8.40 \pm 0.48 $
\end{tabular}
\end{center}
\end{table}

However, the optical pumping is not perfect and we must take this into account for a more realistic estimate of the AC phase.
The interferometer signal (eq. (\ref{I0V})) is the sum of the contributions of the 8 Zeeman-hyperfine sublevels:

\begin{eqnarray} \label{IallMF}
 I&=& I_0 \sum_{F,m_F} P(F,m_F)   \\
  & & \left[1+ {\mathcal V}(F,m_F) \cos{(\varphi_d+ \varphi_{AC}(F,m_F))} \right] \nonumber
\end{eqnarray}

\noindent where $P(F,m_F)$ is the normalized population of the $F,m_F$ sublevel  and, for simplicity, we have omitted all the perturbation phases except the AC phase. The contributions of the  $m_F \ne \pm 2$ sublevels play a minor role for three reasons: their populations are small, the AC phase change sign with the sublevel and the visibility $\mathcal V(F,m_F)$ of their contributions should be small, because the compensator produces a low magnetic field which compensates exactly the Zeeman phase shift only for the $m_F = \pm 2$ sublevels for which the Zeeman effect is purely linear.  As a consequence, we consider that the contributions from the $m_F \ne \pm 2$ sublevels can be neglected and then, a straightforward calculation shows that our experiment measures a corrected AC phase with a weight function of the populations of the populations $P(F=2,m_F=\pm 2)$:

\begin{equation} \label{ACtheoP}
 \varphi^{c}_{AC}(m_F)= \varphi_{AC}(m_F) \frac{P(2,m_F)-P(2,-m_F)}{P(2,m_F)+P(2,-m_F)}
\end{equation}

\noindent  when the optical pumping aims at populating the $F=2, m_F$ sublevel. We thus get more realistic theoretical estimates of the measured AC phases (see Table \ref{table3}) with errors bars which take into account the errors on the population measurements \cite{GillotPO13} and on the capacitor geometry. Our measurements are in  excellent agreement with these corrected theoretical values. Moreover, our results are in very good agreement with an AC phase independent of the atom velocity, as illustrated in Figure \ref{fig5}.

In conclusion, we have measured the Aharanov-Casher (AC) phase by atom interferometry using optically pumped lithium atoms.  Our separated arm interferometer operates with atoms in a single internal quantum state and two opposite electric fields are applied on the interferometer arms to realize the AC measurements. This approach to measure the AC phase is new, as all previous works used either unpolarized neutrons \cite{CimminoPRL89} or Ramsey or Ramsey-Bord\'e interferometers with atoms and molecules \cite{SangsterPRL93,SangsterPRA95,GorlitzPRA95,ZeiskeAPB95,Yanagimachi02}.

Our measurements were performed during the study of the He-McKellar-Wilkens (HMW) phase and we had not optimized the setup for the measurement of the AC phase. In particular, a better control of the residual magnetic field in the interaction region would have simplified the analysis. We have verified that the AC phase depends linearly on the electric field strength. We have performed AC phase measurements for three different mean velocities of the atomic beam ($744$, $1062$ and $1520$ m/s) and these measurements are in good agreement with the fact that this phase does not depend on the atom velocity, a characteristic of a geometric phase. Our measurements, with a statistical uncertainty close to $3$\%,  agree with their theoretical estimates, which have a considerably larger uncertainty  due to their sensitivity to the population distribution over the Zeeman-hyperfine sublevels. The optical pumping of the lithium beam is quite efficient, with about $95\pm 5$\% of the population transferred in the pumped sublevel, but a better pumping, which is feasible \cite{SchinnJOSAB91}, would reduce the uncertainty on the theoretical values of the AC phase.

\acknowledgments
We thank the laboratory technical and administrative staff for their help. We thank H.~Batelaan and A.~Cronin for fruitful discussions, G. Tr\'enec, A.~Miffre and M.~Jacquey for all the work done on our atom interferometer. We thank CNRS INP, ANR (grants ANR-05-BLAN-0094 and ANR-11-BS04-016-01 HIPATI) and R\'egion Midi-Pyr\'en\'ees for support.


\end{document}